**Investigation the Effect of Velocity inlet and Carrying Fluid on the Flow inside Coronary Artery**

Mohammadreza Nezamirad *[1], Nasim Sabetpour [2], Azadeh Yazdi [3], Amirmasoud Hamedi [4]

[1*] PhD student, Department of Chemical, Biological and Materials Engineering, University of South Florida, Tampa, FL, mnezamirad@usf.edu

[2] PhD Candidate, Department of Computer Science, Iowa State University, Ames, Iowa, nasim@iastate.edu

[3] PhD Candidate, Department of civil and environmental engineering, George Mason University, agharibr@gmu.edu

[4] Post Doc, Department of Civil and Environmental Engineering, Florida International University, Miami, FL, USA, amirmasoud.hamedi1@fiu.edu

**Abstract**

In this study OpenFOAM 4.4.2 was used to investigate flow inside the coronary artery of the heart. This step is the first step of our future project, which is to include conjugate heat transfer of the heart with three main coronary arteries. Three different velocities were used as inlet boundary conditions to see the effect of velocity increase on velocity, pressure, and wall shear of coronary artery. Also, three different fluids namely the University of Wisconsin solution, Gelatin, and blood was used to investigate the effect of different fluids on flow inside coronary artery. A code based on Reynolds Stress Navier Stokes (RANS) equations was written and was implemented with the real boundary condition that was calculated based on MRI images. In order to improve the accuracy of the current numerical scheme, hex dominant mesh is utilized. When the inlet velocity increases to 0.5 m/s, velocity, wall shear stress, and pressure increase at the narrower parts.

1. Introduction

Modeling and simulation of heart is a very complex phenomena which can take significant amount of time and research facilities. One of the biggest problems that nowadays donors and recipients are dealing with is the vast geographical differences that can make the process of compatible transplantation very hard and in many cases impossible [1, 2]. In large arteries, shear stress can be exerted linearly [3]. A model for design of cardiovascular devices by acquisition of COMSOL Multiphysics software was introduced [4]. Distention and wall shear stress were used to validate the mentioned model. Non-Newtonian blood flow was simulated by acquisition of laminar module in order to investigate lateral angle (LA) ratio main middle cerebral artery (MCDA) which can be correlated to aneurysm formation. It was found that varying MCDA bifurcation angles can increase aneurism due to wall shear stress and changes the fluid flow. Patient specific modeling for measuring wall shear stress and hemodynamic in coronary arteries with aneurism which was caused by Kawaski Desease (KD) was investigated by Sengupta et al. [5]. In the current study, since our next purpose is heart cooling for transplant purposes, Gelatin and UW solutions was investigated beside blood for flow inside coronary artery.

2. Methodology

The human heart geometry used for simulations was obtained from three-dimensional, high resolution MRI scans. Figure 1 demonstrates different domain surfaces of the heart in different colors. Figure 1 shows a complex model of heart which has three coronary arteries. According to figure 1a, full heart model with three coronary arteries can be seen. Figure 1b shows three coronary arteries which are located in their correct places. The model which is shown in figure 1 was obtained from MRI image from University of Minnesota. Figure 1c shows coronary artery which will be simulated in this study using OpenFOAM 4.4.2.

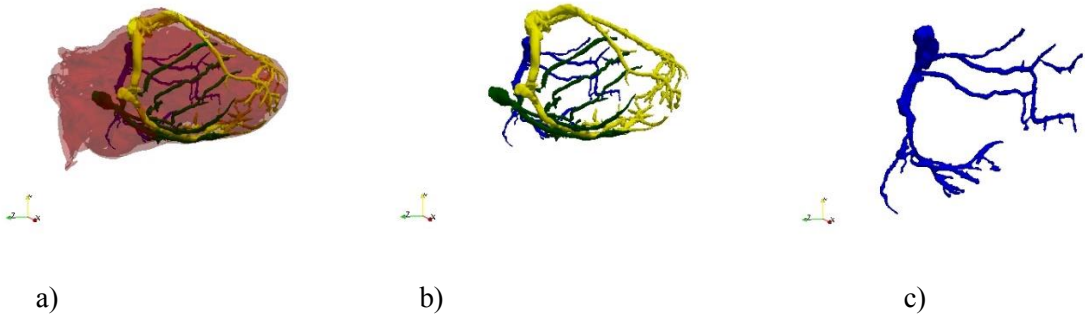

a)                              b)                              c)

Figure 1. The whole heart geometry with three coronary arteries inside it.

Figure 2 shows the mesh which was used for the present study in order to investigate flow inside coronary artery. Figure 2a is the whole geometry of coronary artery. Figure 2b shows close up view for the inlet of coronary artery with its mesh. In this study, three different boundary conditions for three different fluids were used. At first it was assumed that inlet velocity is 0.2 (m/s) and then inlet velocity was changed to 0.35 (m/s) and 0.5 (m/s). In all of the cases outlet velocity was set to zero. Three different fluid that was used in this study are orderly, University of Wisconsin (UW) solution, Gelatin and blood.

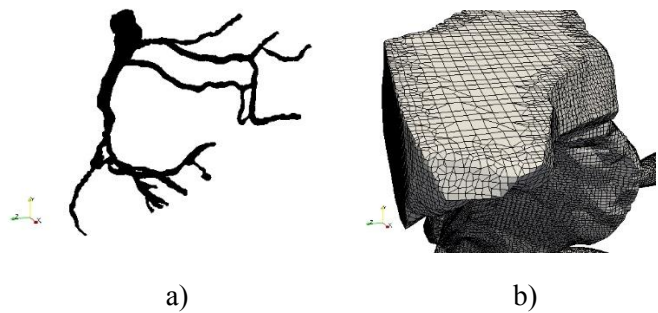

a)                              b)

Figure 2. Mesh used for coronary artery which was used for the present study.

3. Result and discussion

Figure 3 shows pressure distribution when UW solution was used. Figure 3a shows pressure distribution inside coronary artery when inlet velocity is 0.2 m/s and outlet velocity is zero. Figure 3b shows pressure distribution when inlet velocity is 0.35 m/s and outlet velocity is zero. Figure 3c shows velocity distribution when inlet velocity is 0.35 m/s and outlet velocity is zero. According to figure 3, increasing velocity from 0.2 m/s to 0.35 m/s, increases pressure from 8.79 Pa to 20.2 Pa which is nearly 130 percent. By comparing figure 3b and figure 3c, it can be inferred that as velocity increases from 0.35 m/s to 0.5 m/s, pressure increases from 20.2 Pa to 39.2 Pa, which shows 94 percent increase in the value of pressure inside coronary artery.

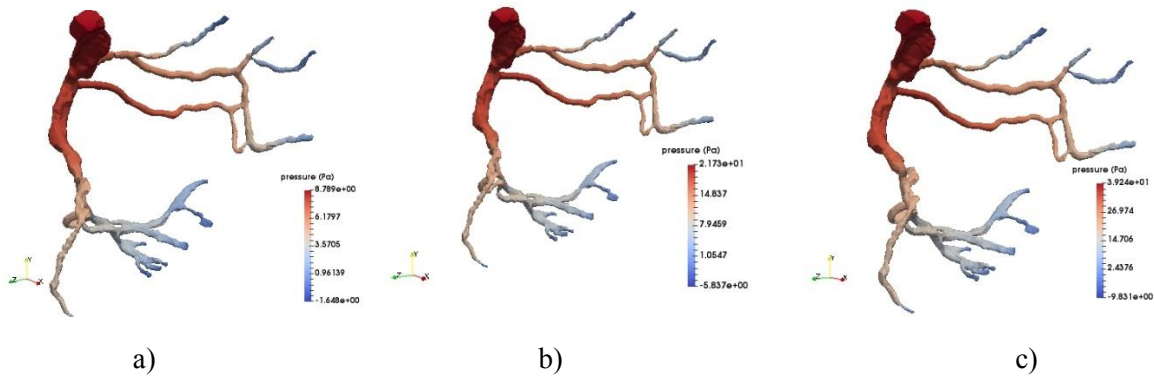

a) b) c)

Figure 3. Pressure distribution when UW solution was used (a. Inlet velocity= 0.2 m/s, b. Inlet velocity= 0.35 m/s, c. Inlet velocity= 0.5 m/s).

Figure 4 shows velocity distribution along coronary artery when the flowing fluid is UW. Three different cases were investigated. In case 1, inlet velocity was 0.2 m/s and outlet velocity were set to zero. In case 2 inlet velocity was 0.35 m/s and outlet velocity were zero m/s. Finally, in case 3, inlet velocity was 0.5 m/s and outlet velocity were zero. From figure 4, it can be inferred that as velocity inlet is 0.2 m/s, the maximum velocity inside coronary artery reaches to 2.49 m/s. As inlet velocity increases to 0.35 m/s, maximum velocity reaches to 4.36 m/s. Finally, when inlet velocity increases to 0.5 m/s, maximum velocity reaches to 6.2 m/s. Maximum velocity in figure 4 can mostly be observed in the regions that diameter of coronary decreases rapidly.

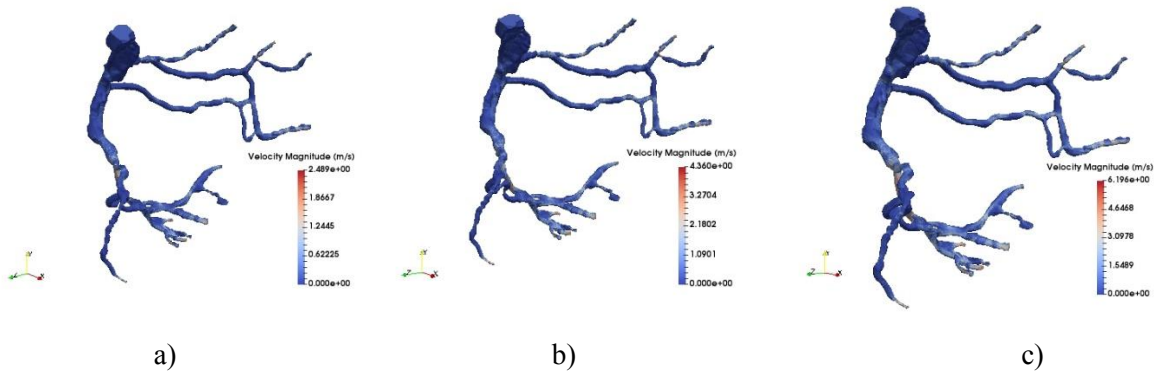

a) b) c)

Figure 4. Velocity distribution when UW solution was used (a. Inlet velocity= 0.2 m/s, b. Inlet velocity= 0.35 m/s, c. Inlet velocity= 0.5 m/s).

Figure 5 shows magnitude of wall shear stress as different inlet velocities were used to evaluate flow field in coronary artery. As figure 5a shows, when inlet velocity is 0.2 m/s, maximum value of wall shear stress mostly forms toward outlet of coronary artery and it reaches to its peak value which is 0.236 ($\frac{m^2}{s^2}$). After changing inlet velocity to 0.35 m/s, wall shear stress inside coronary artery increases to 0.462 ($\frac{m^2}{s^2}$) and it forms mostly in the regions which has smaller diameter. When inlet velocity is 0.5 m/s, figure 5c, wall shear stress increases to 0.69 ($\frac{m^2}{s^2}$) and more regions experience higher amount of wall shear stress compared to other regions.

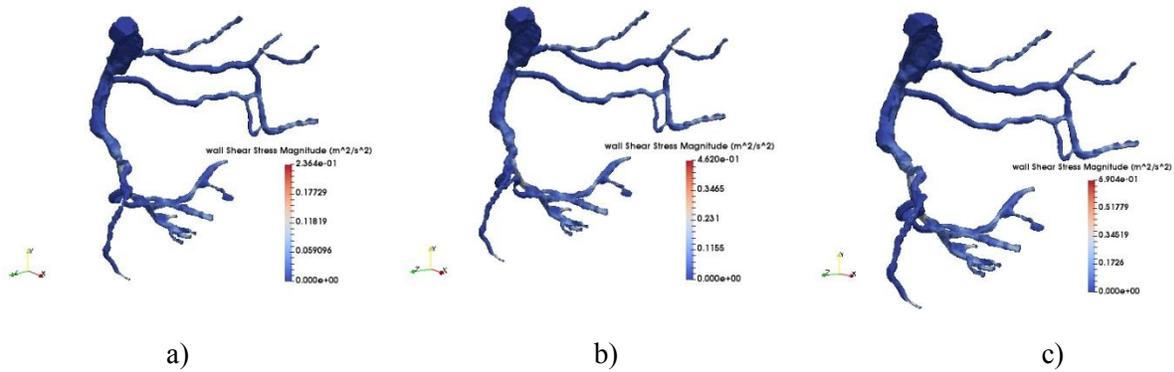

a)  b)  c)

Figure5. Magnitude of wall shear stress for three different cases (case a. Inlet velocity= 0.2 m/s, case b. Inlet velocity= 0.35 m/s, case c. Inlet velocity= 0.5 m/s).

Figure 6 shows pressure inside coronary artery when inlet velocity is 0.2 m/s and outlet velocity is zero. For case a in figure 6, UW solution was used and as mentioned before pressure reaches to 8.79 Pa. By comparing figure 6a, 6b and 6c, it can be found that since Gelatin has the greatest amount of kinematic viscosity ($51e^{-7} \frac{m^2}{s}$), the biggest amount of peak pressure (10.15 Pa) also occurs in this case. Blood due to lowest amount of kinematic viscosity ($28e^{-7} \frac{m^2}{s}$) has the lowest amount of peak pressure (7.96 Pa).

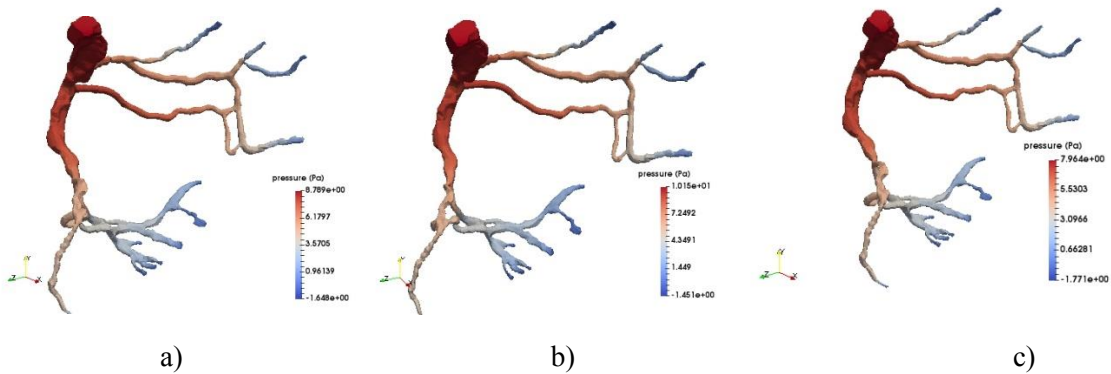

a)  b)  c)

Figure 6. Pressure inside coronary artery when three different solutions was used and inlet velocity was 0.2 m/s (case a. UW solution, case b. Gelatin, case c. blood).

Figure 7 shows velocity distribution inside coronary artery when three different fluids were used. Since difference in kinematic viscosity cannot affect velocity significantly, velocity inside coronary artery as kinematic viscosity changes does not change noticeably.

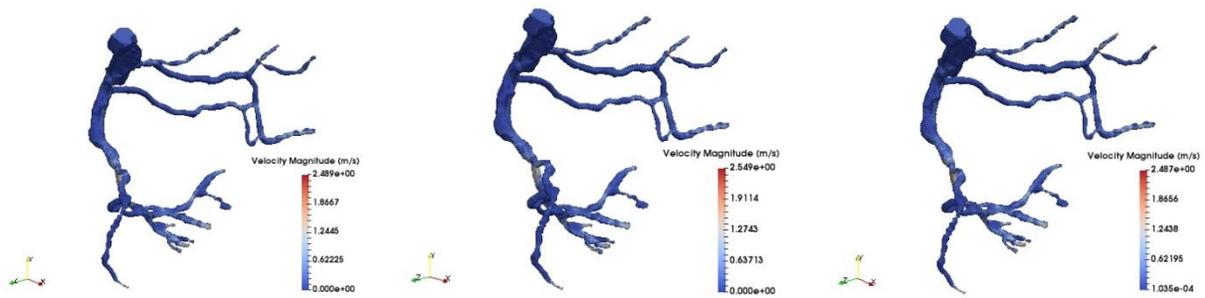

Figure 7. Velocity inside coronary artery when three different solutions were used and inlet velocity was 0.2 m/s (case a. UW solution, case b. Gelatin, case c. blood).

In OpenFOAM, the unit of wall shear stress is $\frac{m^2}{s^2}$. Figure 8 shows distribution of wall shear stress when three different fluids were used. Among the three different cases which are shown in figure 8, case a which is UW solution has the highest amount of wall shear stress but it only occurs in few regions of coronary artery. Although using Gelatin (case b) can be a reason for low value of maximum wall shear stress but in some regions of case b high values of wall shear stress is noticeable.

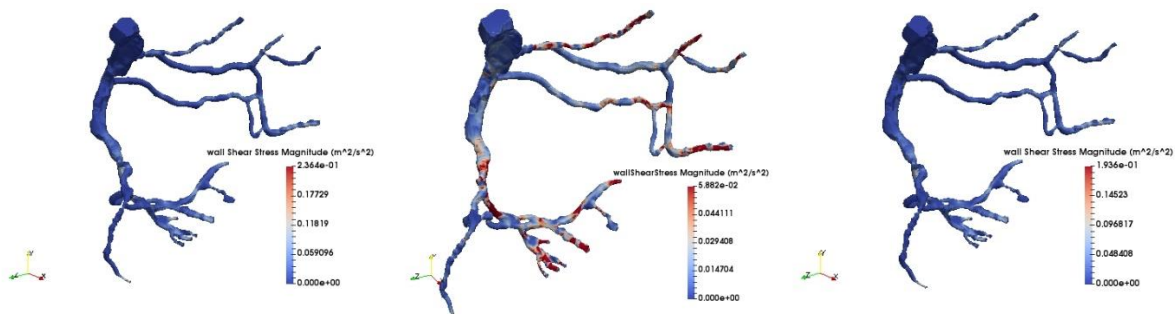

Figure 8. Magnitude of wall shear stress inside coronary artery when three different solutions were used and inlet velocity was 0.2 m/s (case a. UW solution, case b. Gelatin, case c. blood).

4. Conclusion

In this study, OpenFOAM 4.4.2 open source software was used to simulate flow inside coronary artery. SimpleFOAM solver was used to solve Reynolds Average Navier Stoks (RANS) equations. It was found that as inlet velocity increases inside coronary artery, pressure, velocity and wall shear stress increases when the carrying fluid is UW. Also, three different carrying fluids was used in order to investigate the effect of different fluids on velocity, pressure and wall shear stress distribution and it was found that using

Gelatin can cause higher value of peak velocity and pressure. Amazingly the highest value of peak pressure was for UW solution but it was occurred in few regions. Although the lowest value of wall shear stress was for Gelatin, maximum value of wall shear stress occurred in numerous regions.